\RequirePackage{amsmath}
\documentclass[runningheads]{llncs}

\usepackage{graphicx}

\usepackage{float}
\usepackage{multicol}
\usepackage{color}
\usepackage{enumerate}

\usepackage{bbding}

\begin{document}

\title{Neural Review Rating Prediction with Hierarchical Attentions and Latent Factors}
\titlerunning{Rating Prediction with Hierarchical Attentions and Latent Factors}
\authorrunning{Xianchen Wang, Hongtao Liu et al.}

\author{Xianchen Wang\inst{1}\and Hongtao Liu\inst{1}\Envelope{}\and  Peiyi Wang\inst{1}\and Fangzhao Wu\inst{2} \and Hongyan Xu\inst{1} \and Wenjun Wang\inst{1} \and Xing Xie\inst{2}}

\institute{ College of Intelligence and Computing, Tianjin University, Tianjin, China \\ 
\email{ \{wangxc, htliu, wangpeiyi9979, hongyanxu,  wjwang\}@tju.edu.cn} \and 
Microsoft Research Asia, Beijing, China \\ \email{
wufangzhao@gmail.com, xing.xie@microsoft.com
}
}
\maketitle    

\begin{abstract}
Text reviews can provide rich useful semantic information for modeling users and items, which can benefit rating prediction in recommendation.
Different words and reviews may have different informativeness for users or items. Besides, different users and items should be personalized.
Most existing works regard all reviews equally or utilize a general attention mechanism. 
In this paper, we propose a hierarchical attention model fusing latent factor model for rating prediction with reviews, which can focus on important  words and informative reviews. 
Specially, we use the factor vectors of Latent Factor Model to guide the attention network and combine the factor vectors with feature representation learned from reviews to predict the final ratings. 
Experiments on real-world datasets validate the effectiveness of our approach.
\end{abstract}

\section{Introduction}

Using text reviews to model user preferences and item features for rating prediction in recommendation has been an active research topic in recent years~\cite{mcauley2013hidden,diao2014jointly,wang2011collaborative,kim2016convolutional,chen2018neural,lu2018coevolutionary}. 
Kim et al.~\cite{kim2014convolutional} adopt convolutional neural network to extract semantic features of  reviews.
Lu et al.~\cite{lu2018coevolutionary} introduce attention mechanism to build recommender models. 
However, existing works ignore different words in a review and different reviews are differentially informative. 
Besides, they utilize a general attention for all items and users, which may be unreasonable since different users and items should be personalized.

Hence we develop a  Hierarchical Attentions model which incorporates Latent Factor model (HALF) for rating prediction. We utilize hierarchical attention mechanism to focus important words and informative reviews. Specially, we use the factor vectors obtained in Latent Factor Model (LFM) ~\cite{koren2009matrix} as query vectors to guide the review level attention mechanism. 
Moreover, we combine the feature representation learned from text reviews with the factor vectors in LFM to compute the ratings. 
Our experimental results on real-world datasets indicate that HALF considerably outperforms previous methods.
\begin{figure}[H]
\centering
\includegraphics[width=0.9\textwidth]{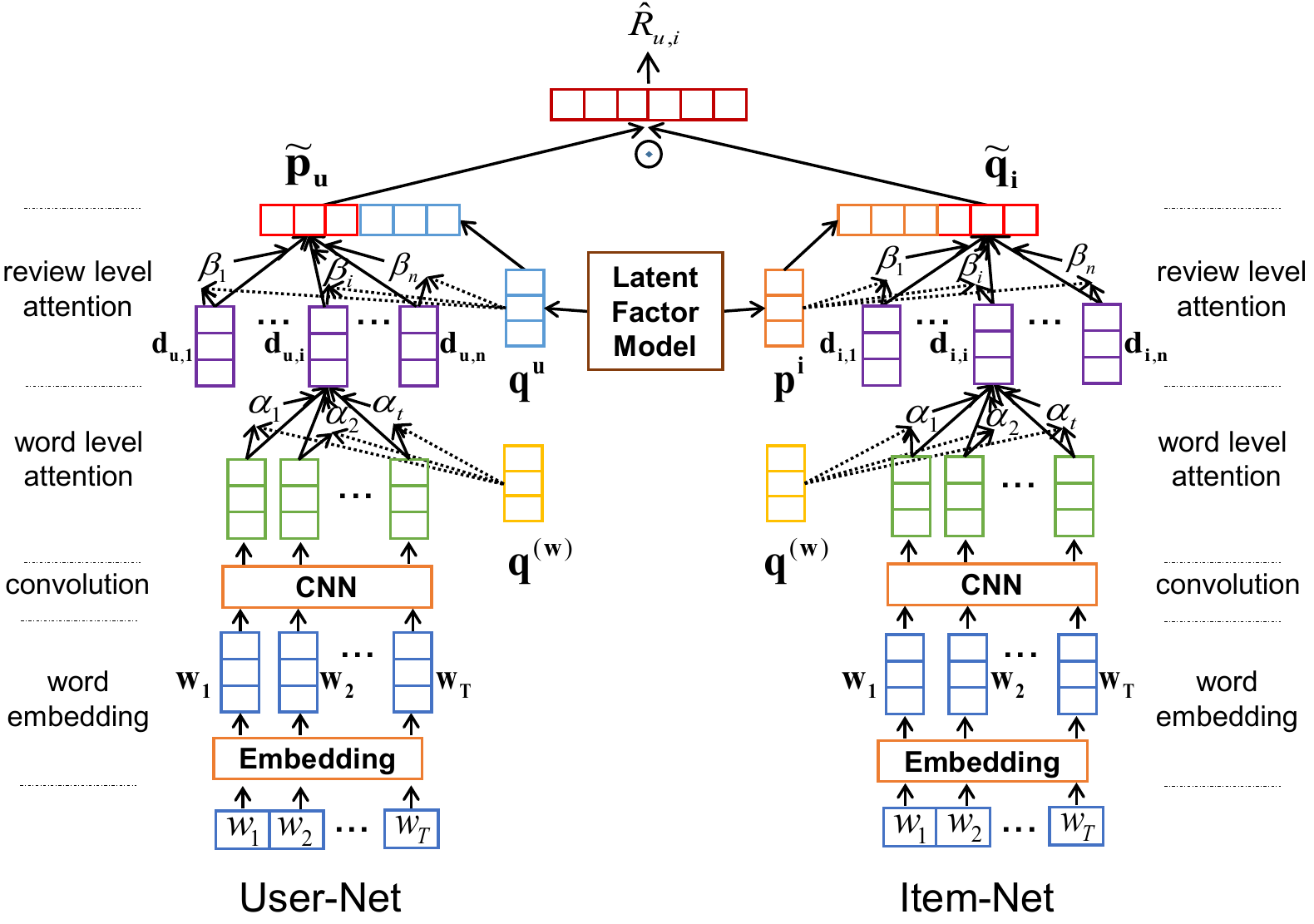}
\caption{\label{fig:overview} Overview architecture of our model.}
\end{figure}
\section{Methodology}
In this section, we will present our model HALF including User-Net and Item-Net. 
Since the architectures of User-Net and Item-Net are similar, we will describe the User-Net in detail only. 
The overview  of our model is shown in Fig.~\ref{fig:overview}.

\subsection{Review Encoder with Word Level Attention}
First, we denote the user set as $\mathit{U}$, item set as $\mathit{I}$, the rating matrix as $\mathbf{R} \in \mathcal{R}^{|U| \times |I|}$ and the text review collection as $D \in \mathcal{R}^{|U| \times |I|}$ and each review is  a word sequence.
For a review $d_{u,i} = \{w_1, \cdots, w_T\}$, we first transform the word sequences into an embedding matrix $\mathbf{M_{u,i}}$ via word embeddings.
Then we apply convolutional neural networks  to extract feature matrix about the review: $\mathbf{C}\in\mathcal{R}^{K \times T}$, and $\mathbf{C_j} = \sigma(\mathbf{W_j * M_{u,i} + b_j}), 1 \leq j \leq K$ where $*$ is the convolution operator, $\mathbf{W_j}$ is the weight matrix of the $j$-th filter and $K$ is the number of filters.

Hence, each column in $\mathbf{C}$ (denoted as $\mathbf{z_i} \in \mathcal{R}^{K}$) represent the semantic feature of the $i$-th word in a review.
To highlight the important words, we employ the attention pooling mechanism in word level, denoted as:
\begin{equation}
    \mathbf{g_i = q^{(w)}Az_i} \ ,
\end{equation}
\begin{equation}
    \alpha_i = \frac{\exp(g_i)}{\sum_{j=1}^T \exp(g_j)}, \ \alpha_i \in (0, 1) \ ,
\end{equation}
where $\mathbf{A}$ and $\mathbf{q^{(w)}}$  are the attention parameters.
Finally, we obtain the representation of the $i$-th review of user $u$ via aggregating feature vectors of all words:
$ \mathbf{d_{u,i}}=\sum_{j=1}^{T}\alpha_j \mathbf{z_j}$.

\subsection{Review Level Attention Guided by Latent Factor Model}
In this section, we employ an attention mechanism in review level  based on Latent Factor Model (LFM)~\cite{koren2009matrix} to focus on more informative and personalized reviews. 
Latent Factor Model predicts the rating $R_{u,i}$ between user $u$ and item $i$ as follows:
$ R_{u,i} = \mathbf{q_u}^{T}\mathbf{p_i} + b_u + b_i + \mu
$
where $b_u, \ b_i$ and $\mu$ are the user bias, item bias and global rating bias.
$\mathbf{q_u}$ and $ \mathbf{p_i}$ are the factor vectors of user $u$ and item $i$ respectively.

Given the review set of a user $d_{u} = \{ {d_{u,1}, d_{u,2}, \cdots, d_{u,N}}\}$, we compute the weight $\beta_j$ about the $j$-th review of the user $u$ as follows:
\begin{equation}
    e_j = \mathbf{q_{u}A_2d_{u,j}} \ ,
\end{equation}
\begin{equation}
    \beta_j = \frac{\exp(e_j)}{\sum_{k=1}^{N}\exp(e_k)}, \ \beta_j \in (0,1) \ ,
\end{equation}
where $A_2$ is the parameter matrix in attention; $\mathbf{q_{u}}$ is the factor vector specifically for the  user $u$.  
The user feature vector is denoted as $\mathbf{m_{u}} = \sum_{i=1}^{N} \beta_j \mathbf{d_{u,i}}$ via aggregating all the review features. Similarly, we can obtain the item features denoted as $\mathbf{m_i}$.

\subsection{Prediction Layer: Fusion of Attention and Latent Factor}
We calculate the rating between a user $u$ and an item $i$ via fusing of attention model and Latent Factor Model as shown in~Fig.~\ref{fig:overview}.
First, we combine the factor vectors in LFM and feature vectors learned from text reviews as the final representation of users and items:
$\tilde{\mathbf{p_u}} = \mathbf{m_u} \oplus \mathbf{q_u}$ 
and 
$\tilde{\mathbf{q_i}} = \mathbf{m_i} \oplus \mathbf{p_i}$
where $\oplus$ is the concatenation operation. 
Afterwards, we compute the final prediction rating that user $u$ would score item $i$ in a form of LFM:

\begin{equation}
    \hat{R}_{u,i} =  \mathtt{ReLU}(\mathbf{W(\tilde{p_u} \odot \tilde{q_i})} + b_u + b_i + \mu)
\end{equation}

\noindent where $\odot$ is the element-wise inner-product operation and $\mathbf{W}$ is the parameter matrix. 
In addition, we employ the mean squared error (MSE) as the loss function.

\section{Experiments}
\textbf{Dataset} 
Following the previous work~\cite{lu2018coevolutionary}, we use five public real-world datasets for evaluation. 
Yelp 2013 and Yelp 2014 are selected from Yelp Dataset Challenge\footnote{https://www.yelp.com/dataset/challenge}. Electronics, Video Games and Gourmet Foods are selected from Amazon 5-core\footnote{http://jmcauley.ucsd.edu/data/amazon/}.
The details of datasets can be found in~\cite{lu2018coevolutionary}.

\noindent{\textbf{Results}} Table~\ref{tab2} shows the results of our model and some recent state-of-art methods in terms of MSE. 
We can conclude that our model HALF can consistently outperform all the baseline methods which indicates the effectiveness of our method.
\begin{table}[H]
\caption{\textbf{Results of our HALF model  and baseline methods in terms of MSE.}}\label{tab2}
\centering
\setlength{\tabcolsep}{1.9mm}{
\begin{tabular}{|c|c|c|c|c|c|}
\hline
  & Yelp 2013 & Yelp 2014 & Electronics & Video Games & Gourmet Foods\\
\hline
JMARS~\cite{diao2014jointly} & 0.970 & 0.998 & 1.244 & 1.133 & 1.114\\
ConvMF+~\cite{kim2016convolutional} & 0.917 & 0.954 & 1.241 & 1.092 & 1.084\\
NARRE~\cite{chen2018neural} & 0.879 & 0.913 & 1.215 & 1.112 & 0.986 \\
TARMF~\cite{lu2018coevolutionary} & \bf 0.875 & 0.909 & 1.147 & 1.043 & 1.019\\
\hline
\bf HALF & \bf 0.875 & \textbf{0.903} & \textbf{1.097} & \textbf{1.016} & \textbf{0.947}\\
\hline
\end{tabular}}
\end{table}

\section{Conclusion}
In this paper, we propose a neural hierarchical personalized attention model HALF which integrates latent factor model into the attention mechanism for rating prediction in recommendation. Experimental results show that HALF significantly outperforms the state-of-the-art models.

\noindent \textbf{Acknowledgments.} This work was supported by the National Social Science Foundation Project(15BTQ056), the National Key R$\&$D Program of China ( 2018YFC0809800, 2016QY15Z2502-02, 2018YFC0831000 ), the National Natural Science Foundation of China (91746205, 91746107), the Key R$\&$D Program of Tianjin ( 18YFZCSF01370 ).

\bibliographystyle{splncs04}
\bibliography{dasfaa19}

\end{document}